\documentclass[structabstract]{aa} 
 \usepackage{longtable}
\usepackage{epsfig}
\usepackage{graphicx}
\usepackage{color}

\usepackage{txfonts}
\usepackage{amssymb}
\usepackage{natbib}                          
\bibpunct{(}{)}{;}{a}{}{,}
\newcommand{\be}{\begin{equation}}
\newcommand{\ee}{\end{equation}}
\newcommand{\bea}{\begin{eqnarray}}
\newcommand{\eea}{\end{eqnarray}}

\begin{document}

 \title{Modeling the EUV spectra of optically thick boundary layers of dwarf novae in outburst}
\author{
V.~Suleimanov\inst{1,2},
M.~Hertfelder\inst{1},
K.~Werner\inst{1},
W.~Kley\inst{1}
}

\offprints{V.~Suleimanov}
\mail{e-mail: suleimanov@astro.uni-tuebingen.de}

\institute{
Institut f\"ur Astronomie und Astrophysik, Kepler Center for Astro and
Particle Physics,
Universit\"at T\"ubingen, Sand 1,
 72076 T\"ubingen, Germany
\and
Kazan (Volga region) Federal University, Kremlevskaja str., 18, Kazan 420008, Russia
}

\date{Received xxx / Accepted xxx}

   \authorrunning{Suleimanov et al.}
   \titlerunning{Spectra of boundary layers}

\abstract
{Disk accretion onto weakly magnetized white dwarfs (WDs) in cataclysmic variables
(CVs) leads to the formation of a boundary layer (BL) between the accretion disk and
the WD, where the accreted matter loses its excess kinetic energy and angular momentum.
It is assumed that angular momentum is effectively transported in the BL, but the transport 
mechanism is still unknown.}
{Here we compute  detailed model spectra of recently published  optically thick one-dimensional
radial  BL models  and
qualitatively compare them with
observed soft X-ray/extreme ultraviolet (EUV) spectra of dwarf novae in outburst.
}
{Every considered BL model with given effective temperature and surface density
radial distribution is divided into a number of rings, and {for each ring, a structure}
model along the vertical direction is computed using the stellar-atmosphere method. The
ring spectra are then combined into a BL spectrum taking  Doppler broadening 
and limb darkening into account.}
{Two sets of model BL spectra are computed, the first of them consists of BL
models with fixed WD mass ($1\,M_{\odot}$) and various relative WD angular
velocities (0.2, 0.4, 0.6 and 0.8 break-up velocities), while the other deals with
a fixed relative angular velocity (0.8 break-up velocity) and various WD masses
(0.8, 1, and 1.2 $M_{\odot}$). The model spectra show broad absorption features
because of blending of numerous absorption lines, and emission-like features at
spectral regions with only a few strong absorption lines. The model spectra are
very similar to observed soft X-ray/EUV spectra of SS~Cyg and U~Gem in
outburst. The observed SS~Cyg spectrum could be fitted by BL model spectra with 
WD masses 0.8 - 1 $M_{\odot}$ and relative angular velocities 0.6  - 0.8 break up velocities.
These BL models  also  reproduce the observed ratio
of BL luminosity and disk luminosity. The difference between the observed and the BL model
spectra is  similar to a hot optically thin plasma spectrum and could be associated with the
spectrum of outflowing plasma with a mass loss rate compatible with the BL mass accretion rate.}
{The suggested method of computing BL spectra  seems very promising and
{can} be applied to other BL models for comparison with EUV spectra of dwarf novae in outburst.}


\maketitle
%

\section{Introduction}

The importance of energy release between an accretion disk and a central object
with a surface was realized almost immediately \citep{LBP:74} after the
introduction of modern accretion disk theory \citep{ShS:73}. Later a similar
one-dimensional (1D) boundary-layer (BL) theory was developed 
 \citep{Pringle:77, PS:79, Tylenda:81, Reg:83, BR:92, RB:95, GRS:95,  PNr:95}.
 Here we mainly consider BLs
around white dwarfs \citep[cataclysmic variable stars (CVs), see review
in][]{BW:03} and describe a few key points of BL theory.

Depending on the accretion rate, $\dot M$, a BL can be optically thin ($\dot M <
10^{16}$~g s$^{-1}$) or optically thick ($\dot M > 10^{16}$~g s$^{-1}$)
\citep{PS:79}. 
 The BL optical thickness also depends on the mass and angular velocity of the white dwarf (WD),
as well as   on the value of turbulent viscosity in the BL \citep{PNr:95, Coll.etal:00}.
For the description of optically thick BLs two qualitatively
different approaches were suggested, which reduced the problem to a 1D model. 
In the first approach, the BL is considered as the inner part of the
1D  axi-symmetric accretion disk \citep{PS:79, Reg:83, PNr:95} with no vertical component of the  
velocity. The condition  imposed on the BL at the inner boundary is that it rotates at the stellar equatorial velocity.
Therefore, the accreting matter has to decelerate and  release its excess
energy in a relatively narrow (a few percent of the inner disk radius) ring.
Most of this energy is radiated away, but part of it can be advected into the WD \citep{Popham:97, Godon:97}, 
deposited to the outflow \citep{MR:00}, and can accelerate the outer layers of the WD
forming a fast rotating belt \citep{Long.etal:06}. 
In contrast, in the second approach, the matter keeps its nearly Keplerian  velocity at the central object's
equator and spreads over the surface \citep{IS:99}. The
spreading matter loses its kinetic energy gradually due to friction with the more
slowly rotating surface of the central star and radiates the released energy
mainly in two bright high-latitude belts. Initially the spreading layer model
was developed for BLs around neutron stars \citep{IS:99}, but later it was
extended to the white dwarf case, too \citep{PB:04}.

Together with the difference in geometry, the description of the viscosity,
which provides a strong coupling of rings of matter moving with
different velocities and an effective thermal dissipation of the lost energy, is
different in both approaches. In the first model the usual $\alpha$ prescription
for accretion disks is used, which means that the $r\varphi$-component of the viscosity
stress tensor is parameterized by the total pressure $P$ at a
given point as $w_{r\varphi}=\alpha P$ \citep{ShS:73} or, almost equivalently, a
similar parametrization of kinematic viscosity, $\nu_\alpha$, is used. In the
second approach the $\alpha$ viscosity is completely ignored and only the
friction between the high-velocity spreading matter and the dense and relatively
cool stellar envelope is considered. The corresponding component of the specific
frictional force, $f_{\rm sl}= \alpha_{\rm b} \,\rho_{\rm b} {\rm v}_{\rm sl}^2$,
is again scaled using the matter density at the bottom of the spreading layer,
$\rho_{\rm b}$, and the relative spreading matter velocity, ${\rm v}_{\rm sl}$
\citep{IS:99}. The estimated  value of $\alpha_{\rm b}$ is relatively low, about 10$^{-3}$
\citep{IS:99}.  Detailed consideration of this problem shows that the matter
deceleration due to friction with the underlying stellar envelope is not trivial
and has to be investigated more carefully \citep{IS:10}.

It is clear from both viscosity descriptions that we do not know the physics of
kinetic energy loss and angular momentum transfer in BLs \citep[see,
e.g.,][]{KI:94, Narayan.etal:94, Godon:95}. The magneto-rotational instability
(MRI) \citep{Vel:59, Chandra:60, BH:91}, which is usually considered as a physical realization of
the $\alpha$ viscosity, cannot operate in the case of angular velocity decrease
taking place in BLs \citep{Godon:95}. Recently, \citet{BR:12} and \citet{BRS:12,BRS:13}
have suggested a new physical model for an angular momentum transport in BLs based on
acoustic instabilities. Various hydrodynamical instabilities,
beginning with simple shear instabilities \citep{KT:78}, were also considered
before as a way of angular momentum transport \citep[see the review
in][]{BR:12}.

Another important feature of BLs is their  3D nature. A full 3D
treatment is important to account for a correct turbulent viscosity description. But BL models that 
use  a parametrization of the turbulent viscosity  could be considered as axisymmetric 2D models. The first
attempts to describe the BL around the WD using a 2D time-dependent
hydrodynamical approach were performed many years ago \citep{RF:86, KH:87,
Kley:89}. The importance of the viscosity prescription was also demonstrated
\citep{Kley91}. Recently, this kind of computation was repeated with higher,
albeit  insufficient temporal and spatial resolution without \citep{FB:05} and with
rudimentary radiation treatment \citep{Balsara.etal:09}. These investigations have
confirmed that matter is spreading over the WD surface for optically thick BLs.

Astrophysical observations give the possibility to constrain the correct
description of angular moment transport and energy dissipation in BLs. It is
necessary to compare properties of a mature BL model with observed
features. There are two ways for this kind of comparison. The first one is to
investigate  flux variability, which is probably connected with BLs, using
 a noise power spectrum approach, for instance \citet{vdK:89}. There are a lot of
observational data about rapid flux variability in low-mass X-ray binaries
(LMXBs),  such as quasi-periodic oscillations (QPOs) and  power density spectra \citep{vdK:00}. 
The
cumulative data about rapid variability of CVs, especially in the optical band, is
even larger \citep[see, e.g.,][ and references therein]{Patterson:81, W:86, BW:03, WP:08}. 
Two types of rapid oscillations were distinguished \citep{RN:79}, namely high-degree coherent oscillations with
relatively short periods ($\sim$ 7-70 s) in dwarf nova outbursts (DNOs), and less coherent quasi-periodical oscillations (QPOs) with
longer periods (up to tens of minutes). The amplitudes of variability for  both DNOs and QPOs
are relatively low in the optical band ($<$ 0.01 mag). On the other hand, the amplitude of these oscillations
 in X-rays can be much greater \citep[tens of percent,][]{CM:84, JW:92}. It is interesting that the coherent oscillations in SS Cyg  are observed
 in soft X-rays, but not in hard X-rays  \citep{Swank:79, JW:92}. This fact supports the hypothesis
 that the coherent oscillations might be connected with an optically thick BL that could be responsible for the soft X-ray
 radiation of SS Cyg during outburst.  There are a few relatively simple models
to explain DNOs and QPOs \citep[see, e.g.,][]{Popham:99, WW:02, PB:04b, Godon:95b, Coll.etal:98, Coll.etal:00b}, and
part of them are connected with the existing BL models.

The second way is a comparison of observed emergent spectra of close binary
systems with predicted spectra of accretion disks and BLs.  Emergent spectra
could be computed for BL models with the simple local $\alpha$-viscosity, because
those models consider energy release and predict some bolometric radiation flux distribution
over the BL. BL model spectra in blackbody approximation were computed by many
authors \citep[see, e.g. ][]{Tyl:77, Kley91, PNr:95}. The contribution of BL radiation to the FUV spectra of
some CVs was also taken into account by computing models of BL rings using
the stellar-atmosphere method  \citep{GS:11, GS.etal:12}.

Luminous LMXBs (with $L > 0.05-0.1\,L_{\rm Edd}$)  with neutron stars show relatively soft and wide
X-ray spectra, which can be represented to first approximation by two
blackbodies with temperatures $kT$ about 1 keV and 2.5 keV
\citep{Mitsuda.etal:84}. It is possible to assume that the first component
corresponds to the accretion disk, and the second one corresponds to the
BL. Study of X-ray variability helped to distinguish between these two
components \citep{GRM:03, RG:06}. These authors extracted the spectra of the
strongly variable component of the observed X-ray flux in a few LMXBs and found
that they are very similar for all the investigated sources and that
they can be described by a comptonized plasma spectrum with $kT \approx
2.4$\,keV and electron scattering optical depth $\tau_e \approx 5-7$. They
connected these high-variability components of the X-ray flux to the BL
emission. The spectra of these components were described by model spectra of
spreading layers \citep{Sul.Pout:06}. Another reference to the validity of the
hypothesis that the BLs around neutron stars are spreading layers was presented
by \citet{RSP:13}. The high-variability spectral component of the X-ray
transient source XTE J1701-462 keeps the same spectral shape when the bolometric
luminosity varies by a factor of twenty. The maximum temperature of this
component (Wien tail color temperature) is about 2.4 -- 2.6~keV and coincides
with the maximum color temperature of the type~I X-ray bursts of the same
source. It is believed that the latest color temperature of a burst corresponds
to the Eddington luminosity \citep[see, e.g., review][]{LvPT93}. Therefore, most
probably, the BLs in LMXBs consist of a part of the neutron star surface that
radiates at the Eddington limit with the emitting area being proportional to the
luminosity, as it was predicted by the spreading layer model \citep{IS:99}.

X-ray and EUV radiation of CVs, which can be associated with BL emission, is
much more diverse and a univocal interpretation is difficult \citep[see,
e.g.,][]{PR:85b,PR:85}. Blackbody approximation of the optically thick BL model
spectra predicts a luminous soft X-ray / extreme UV component  with temperature
200 -- 500~kK \citep{PS:79, PNr:95, Coll.etal:00, PB:04,HKSW:13} in the spectra of all high mass-accretion rate CVs
without significant magnetic field (dwarf novae in outburst and nova-like
stars). In fact, only a few of them have this kind of component in their
spectra, e.g. SS Cyg and U Gem in outburst \citep{CM:84}. The X-ray spectra of
other non-magnetic CVs are rather hard 
 \citep{vtv:94, vt.etal:96} and can be described by the model of a
cooling flow with temperatures $kT$ from a few keV up to tens of keV
\citep{Done.etal:95, Mukai.etal:03, Baskill:05}. This kind of radiation is
natural for optically thin BLs \citep{PS:79}, or even for the inner hot accretion
flow \citep{Revnivtsev.etal:12} to be expected for dwarf novae in quiescence, but requires some
non-trivial explanation for high mass-accreting CVs, like V603~Aql \citep[see,
e.g. ][]{PR:85}.  Formally, \citet{PNr:95} have found that the transition to the optically thin
regime in the model BLs could be reached at sufficiently high mass accretion rates.
But their results are questionable because they underestimated 
the Rosseland "true" opacity by at least two orders of magnitude, using 
Kramers opacity $k_{\rm kr} = k_0\, \rho T^{-3.5}$ cm$^2$ g$^{-1}$ 
with $k_0 = 6.6 \times 10^{22}$. The more realistic
coefficient is much larger,  $k_0 = 5 \times 10^{24}$ \citep{FKR:02}. 
 
At present time, soft X-ray/EUV components were found in four dwarf novae in
(super-) outbursts: SS Cyg \citep{Cordova.etal:80b, CM:84, Mauche:95, Mauche04}, U Gem
\citep{Cordova.etal:80a, CM:84, Longetal:96}, VW Hyi \citep{Mauche:96a}, and OY Car
\citep{MR:00}. The high-inclination system OY Car shows mainly broad emission
lines arising due to resonance scattering in the strong disk wind
\citep{MR:00}. The high-resolution spectra of the other CVs, obtained by {\sl
EUVE} and {\sl Chandra} observations, show numerous broad absorption- and
emission-like details. These spectra are similar to soft X-ray spectra of
supersoft X-ray sources \citep{Lanz.etal:05, Rauch.etal:10} and also have to be
modeled using stellar model-atmosphere methods. The first attempt to fit the
soft X-ray spectrum of SS Cyg in outburst \citep{Mauche04} by hot LTE
model-atmosphere spectra showed the potential for success of this approach
\citep{SMZW:13}.

With the work at hand, we start to model the various BL models using the
stellar-atmosphere method and a comparison of the results with the properties of
observed soft X-ray/EUV spectra of CVs mentioned above. Here we present the
model spectra of particular 1D hydrodynamic BL models that were computed recently by
\citet{HKSW:13}.
 
\section{Method}
 
Our work is based on the 1D models of BLs between optically thick accretion
disks and WDs that were computed in our previous work \citep{HKSW:13}. Models
are considered in a cylindrical coordinate system ($z$, $\varphi$, $R$), being
axisymmetric (independent of $\varphi$) and vertically averaged (over
$z$-coordinate). They were computed for a fixed mass-accretion rate $\dot M =
1.5 \times 10^{-8}$ $M_\odot$\,yr$^{-1}$, three different values of WD mass,
$M_{\rm WD}$ = 0.8, 1, and 1.2 $M_\odot$, and five values of WD angular
velocity, $\omega_{\rm WD}$ = 0, 0.2, 0.4, 0.6 and 0.8 $\omega_{\rm K}$, where
$\omega_{\rm K}$ is the Kepler angular velocity $\omega_{\rm K}^2 = GM_{\rm
WD}/R^3$ at the WD radius $R_{\rm WD}$. WD radii were calculated using the
\citet{Nauenberg:72} relation. An increase of equatorial radii due to rotation was
ignored. A simple $\alpha$ prescription for the viscosity $\nu_\alpha$ was used
according to \citet{ShS:73}, see details in \citet{HKSW:13}:
\be 
\nu_\alpha = \alpha a^2 \omega_{\rm K}^{-1}, 
\ee
where $a$ is the sound speed. We note that $\omega_{\rm K}$ is taken for the
current BL radius, however, the radial extension of model BLs is low and
$\omega_{\rm K}$ for a given model is almost constant. A relatively low value
of the $\alpha$ parameter ($\alpha=0.01$) is assumed for all models in order to
avoid supersonic radial motions.
 
For every 1D model we consider the distribution of the following physical
parameters along the radial coordinate $R$: effective temperature $T_{\rm eff}
(R)$ (or, equivalently, the total radiated flux $F_0(R)$), surface density
$\Sigma_0(R)$, BL half-thickness $H_0(R)$, and rotation velocity $\rm v_{\rm
\varphi}(R)$. The following steps have to be performed to compute the model
emergent spectrum:
\begin{itemize}
 \item Divide the BL model in a number of rings with equal luminosity.
 \item Compute a gray model for each ring along the vertical coordinate $z$.
 \item Starting from the gray model, compute a model of the each ring using
model atmosphere methods together with the local emergent spectrum.
 \item Sum up the local spectra to a total BL spectrum taking  the
rotation of the rings into account.
\end{itemize}
This approach is almost identical to the one that was used for the computation of CV accretion disk spectra
by many authors \citep{KH:86, SW:91, SV:92, WH:98, Nagel.etal:04}.
We understand that  the local 1D  method assumed here is rather crude. But we believe that it gives 
a reasonable first approximation, which is applicable to comparison with observations. This method is a
first step to more sophisticated 2D models.

Accretion disk and BL in the 1D models presented by \citet{HKSW:13} are
considered as a comprehensive structure, and the boundary between accretion disk
and BL is blurred. However, the $T_{\rm eff}(R)$ distributions have a local
minimum at $R\approx 1.1 R_{\rm WD}$. We adopted that radius as the boundary
between the disk and the BL and studied the BLs at radii lesser than this
boundary. We divided each BL model into a few (5 -- 20) rings, 
and the particular radius $R_j$ of a given ring with number $j$,
for which a local vertical structure and local emergent spectrum are computed,
is determined by the condition 
\be 
F_0(R_j) \Delta R_j = F_0(R_j) \,( R_j^U - R_j^L) = \int_{R^L_j}^{R_j^U}F_0(R)\, dR, 
\ee 
where $R_j^U$ and $R_j^L$ are the radii of the upper and lower ring boundaries,
respectively. We note that $R_{j-1}^U =R_j^L$.

We now describe the other steps of our method in more detail.

\subsection{BL vertical structure. Gray atmosphere approach.}

The approach employed here is based on the numerical method used for modeling of
accretion disk structures over the $z$ coordinate as presented by
\citet{SLS:07},  \citep[see also][]{KH:86, SW:91, SV:92, WH:98, Nagel.etal:04}.  Some
modifications are made and described additionally.
 
The vertical structure of each BL ring at radius $R_j$ is determined by the
ring parameters $T_{\rm eff}$, $\Sigma_0$, $H_0$, and a set of differential
equations. The first one is the hydrostatic equilibrium equation
\be
 \frac{1}{\rho}\frac{dP}{dz'} = g_{\rm z} = \frac{(z_1 -z')}{(1-(z_1 -z')^2/R_j^2)^{1/2}}\,\, 
 \omega^2_{\rm K}(R_j) \, ,
\label{u1}
\ee
where $P = P_{\rm g} + P_{\rm rad}$ is the total pressure, the sum of gas and
radiation pressure, $\rho$ is the matter density, $z_1$ is a parameter, the
distance from the mid-plane to the highest point of the vertical model, and
$z'=z_1-z$ is the vertical coordinate, which is equal 0 at the highest point of
the model.

The second equation is the energy-conservation law. The correct local energy
generation rate $dF/dz$ in accretion disks as well as in BLs is not known and a
local version of the $\alpha$ approach is often used \citep[see
e.g. ][]{SLS:07}. Fortunately, local emergent spectra of optically thick
accretion disks (and, therefore, the BL models considered here, too) are weakly
dependent on details of the energy generation rate, so we took the simplest
version \citep{ShS:73}:
\be
 \frac{1}{\rho}\frac{dF}{dz'} = -2\frac{F_0}{\Sigma_0}.
\label{u2}
\ee
Here $F$ is the integral (bolometric) vertical flux at given height,
$F_0=\sigma_{\rm SB} T^4_{\rm eff}$, is the emergent integral flux. The energy
conservation law has the integral  \citep{ShS:73}
\be
 F(m)=F_0\left(1-\frac{2m}{\Sigma_0}\right),
\label{u3}
\ee
where the boundary condition $F(m=0)=F_0$ is used. Here the Lagrangian coordinate $m$ is determined by the equation
\be
dm = \rho dz',
\label{u4}
\ee
 and, therefore, 
\be
 \Sigma_0 = \int_{-\infty}^{+\infty} dm \approx 2\int_0^{\,z_1} \rho dz'.
\ee
 We assumed a purely radiative transport of energy in $z$-direction, therefore, the
third equation is the radiation transfer equation. For the gray approach we use
the first moment of that equation
\be
\label{u5}
 \frac{1}{\rho}\frac{dP_{\rm rad}}{dz'} = \frac{\kappa_\mathrm{R} F}{c},
\ee
where $\kappa_{\rm R}$ is the Rosseland opacity, which is determined as the
greater value of electron scattering, $\sigma_{\rm e} = 0.335$ cm$^2$ g$^{-1}$,
and Kramers opacity, $\kappa_{\rm kr}= 5\times 10^{24}\rho T^{-3.5}$. The sum of these
values is also relevant \citep{PNr:95}.

These equations are solved together with the ideal gas law
\be \label{igl}
  P_{\rm g} = nkT,
\ee
where $n$ is the total number density of particles. We assume full local
thermodynamic equilibrium (LTE). Therefore, the local gas temperature $T$
can be found from the integrated (over frequency) mean intensity $J$
\be
 J=\frac{3c\, P_\mathrm{rad}}{4\pi}= B(T) =\frac{\sigma_{\rm SB}T^4}{\pi}.
\label{integ_Pl}
\ee
Here we used the following TE relations between radiation energy density
$\varepsilon_{\rm rad}$, integral mean intensity $J$, and radiation pressure:
\be
 \varepsilon_{\rm rad}=\frac{4\pi J}{c} = 3 P_{\rm rad}. 
\ee
A solar chemical composition is assumed and all ionization states of the 15
most abundant chemical elements are taken into account. The relation between
density $\rho$, total number density $n$, and electron number density $n_{\rm
e}$ is found using Saha's equation for each considered ionization state.

\begin{figure}
\centering
\includegraphics[angle=0,scale=1.]{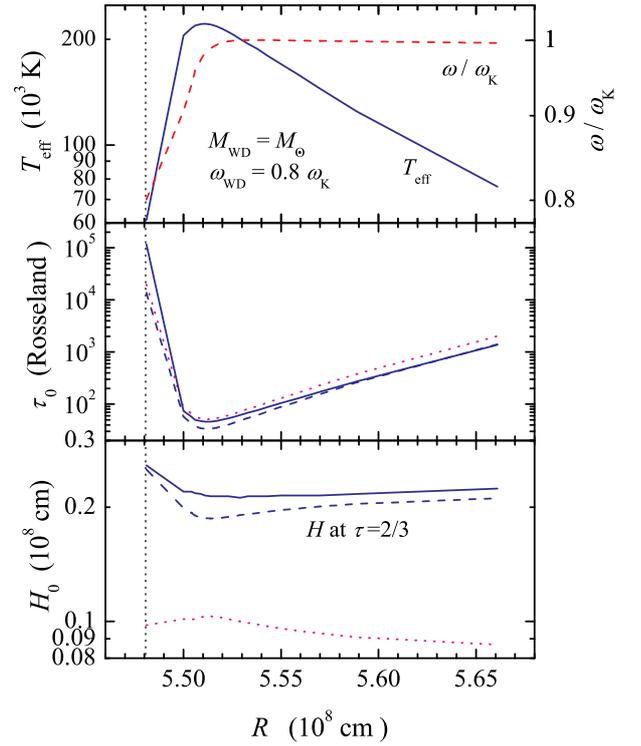}
\caption{\label{fig1}
The distribution of BL quantities along radius.
The data are from the model with $M_{\rm WD} = M_\odot$\, and
$\omega_{\rm WD} = 0.8\,\omega_{\rm K}$. The position of the WD surface
 at $R_{WD}$ = 5.48 $\times$ 10$^8$ cm is shown by the vertical dotted line.
Top panel:
Shown is the effective temperature  (solid curve) 
together with the relative angular velocity (dashed curve),
as obtained from the 1D hydrodynamical model. These are the same for the
atmosphere models.
 Middle and bottom panel:
Rosseland optical half-thickness, $\tau_0$, and geometrical half-thickness, 
$H_0$, for the model-atmosphere models (solid lines). The distributions
obtained from the 1D  hydrodynamical model (dotted curves) and the gray model
(dashed curves) are also shown.
}
\end{figure}

\begin{figure}
\centering
\includegraphics[angle=0,scale=1.]{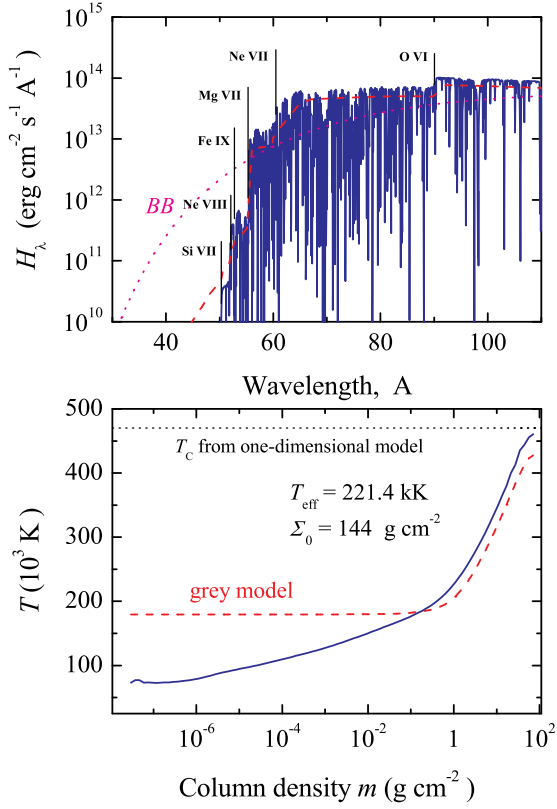}
\caption{\label{fig2}
Top panel: Local spectra of the hottest ring of the same BL model as in
Fig.\,\ref{fig1} computed with (solid curve) and without (dashed curve)
spectral lines taken into account. The blackbody spectrum corresponding to
the effective temperature is shown by the dotted curve. Bottom panel:
Temperature structures of the model with lines (solid curve) and the gray
model (dashed curve). The central temperature of the hydrodynamical 1D model is
shown by the dotted line.}
\end{figure}

We solve Eqs.\,(\ref{u1}) and (\ref{u3}-\ref{u5}) from the surface $z'=0$
to the mid-plane $z'=z_1$ by using a shooting method with the boundary
conditions
\be
 F(0)=F_0, \qquad \rho(0)=0, \qquad
P(0)=P_{\rm rad}(0)= \frac{2}{{3}}\frac{F_0}{c}. 
\ee
There is one additional parameter unknown at the outset: $z_1$. We find it by the
dichotomy method on the range (1 -- 30) $H_0$ using an additional boundary
condition at the mid-plane
\be
 F(z'=z_1) = 0 
\ee
or, equivalently, 
\be
 m(z'=z_1)=\frac{\Sigma_0}{2}.
\ee
A comparison of optical depths and half-thicknesses of the 1D and the
gray-approach models along $z$ coordinate for one particular BL model ($M_{\rm
WD}=M_\odot$\, and $\omega_{\rm WD}=0.8\,\omega_{\rm K}$) is shown in Fig.\,\ref{fig1} 
in the two bottom panels. 
It is clear that the new half-thickness of the BL model
$H=z_1-z(\tau_{\rm R}=2/3)$ is approximately 
twice as large  as the thickness of the 1D hydrodynamic model (red dotted line).
 This is a well
known factor \citep{ShS:73} because the half-thickness of 1D models is just the
pressure scale height $H_0=a \omega_{\rm K}^{-1}$,  while the new thickness is given by the height of the
photosphere. 
The effective temperatures are plotted in the top panel of Fig.\,\ref{fig1},
here the models agree by definition. The gray temperature
structure of the hottest ring of the same BL model is shown in Fig.\,\ref{fig2}
(bottom panel). We note that this ring is a reference model for our illustration
of the presented method.  The parameters of the rings for all computed BL models with five rings are shown in
Tables \ref{tabl1}-\ref{tabl3}. The ring half-thicknesses, $H$, and the ring Rosseland optical depths,
$\tau_0 = \int_0^{z_1} k_{\rm R}\, \rho\, dz$, were computed by using the gray ring models.

\begin{table}[htbp]
\begin{center}
\caption{Parameters of the rings for the models with parameters $M_{\rm WD}~=~M_\odot$ and $\omega_{\rm WD}$ = 0.8, 0.6, 0.4, 0.2\, 
$\omega_{\rm K}$.
}\label{tabl1}
{
\begin{tabular}{llllll}
$R$, 10$^8$ cm & $T_{\rm eff}$, 10$^3$ K & $\Sigma_0$, g/cm$^{2}$ & $H, $ 10$^7$ cm & $\tau_{\rm 0}$ & $\omega / \omega_{\rm K}$      \\ 
\hline
5.493  &	154.4 &	499.6 &	2.188   &   230  &	0.839\\
5.513 &	219.6   &   140.4 &	1.866  &	34.1  &	0.987\\
5.522 &	209.4 &	152.9 &	1.882   &   40.3  &	0.998\\
5.543 &	180.3 &	207.1 &	1.945  &	71.1  &	1.000\\
5.604 &	109.5 &	466.8  &	2.059  &	404  &	0.998\\
&&&&&\\
5.482 &	241.2 &	376	 &	2.36	   &     79.8 &	0.652\\
5.505 &	298.0 &	83.4  &	1.938 &	13.9 &	0.999\\
5.515 &	283.2 &	93.5	 &	1.889 &	15.7 &	1.003\\
5.534 &	253.8 &	121.1 &	1.905 &	23.1	    &    1.002\\
5.610  &	144.1 &	363.1 &	2.094 &	187 &	0.999\\
&&&&&\\
5.479 &	302.6 &	306	   &	2.565 &	51.2&	0.475\\
5.499 &	356.6 &	68.3 &	2.395 &	11.4	 &	0.994\\
5.512 &	333.1 &	75.7 &	2.205 &	12.7 &	1.005\\
5.538 &	302.9 &	101 &	2.089 &	16.9 &	1.004\\
5.623 &	170.1 &	330 &	2.145 &	125	 &	0.999\\
&&&&&\\
5.478 &	349.7 &	264 &	2.998 &   44.1 &	0.299\\
5.498 &	410.0 &	62  &	3.115 &	10.3 &	0.992\\
5.515 &	372.7 &	69.3 &	2.619 &	11.6 &	1.006\\
5.540   &	331.5 &	85.2 &	2.272 &	14.2 &	1.004\\
5.708 &	188.9 &	648 &	2.554 &	203	 &	0.997\\
\hline
\end{tabular}
}
\end{center}
\end{table}

\begin{table}[htbp]
\begin{center}
\caption{Parameters of~ the rings~ for~ the~ models~ with~ parameters  $\omega_{\rm WD} = 0.8\, \omega_{\rm K}$ and 
$M_{\rm WD}$ = 1.2 and 0.8 $M_\odot$.
}\label{tabl2}
{
\begin{tabular}{llllll}
$R$, 10$^8$ cm & $T_{\rm eff}$, 10$^3$ K & $\Sigma_0$, g/cm$^{2}$ & $H, $ 10$^7$ cm & $\tau_{0}$ & $\omega / \omega_{\rm K}$      \\ 
\hline
 3.865 &  222.2  &   565.6 &   1.370  &  179 &    0.829\\
 3.873  & 304.3  &   164.7  &  1.187  &   30.6  & 0.961\\
 3.886  & 284.1  &   176.7   &  1.180   &  36.5  &0.993\\
 3.895  & 253.0  &   227.2   &  1.204  &   56.9  & 0.993\\
 3.936  & 150.5  &   559.9  &  1.275  &   356 & 0.991\\
&&&&&\\
7.044 & 113.8 & 461.3 & 3.166 & 299     & 0.848 \\
7.065 & 167.9 & 147.7 & 2.784 & 46.2 & 0.965 \\
7.092 & 156.8 & 156.0 & 2.788 & 55.1 & 1.002 \\
7.114 & 139.5 & 192.0 & 2.850 & 83.9 & 1.004 \\
7.205 & 84.37 & 426.8 &  3.026 & 482 & 1.001 \\
\hline
\end{tabular}
}
\end{center}
\end{table}

\begin{table}[htbp]
\begin{center}
\caption{Parameters of the rings for the model with parameters $M_{\rm WD}~=~0.8 M_\odot$ and $\omega_{\rm WD} =  0.6\, \omega_{\rm K}$.
}\label{tabl3}
{
\begin{tabular}{llllll}
$R$, 10$^8$ cm & $T_{\rm eff}$, 10$^3$ K & $\Sigma_0$, g/cm$^{2}$ & $H, $ 10$^7$ cm & $\tau_{\rm 0}$ & $\omega / \omega_{\rm K}$      \\ 
\hline
7.022 & 177.7 & 297 & 3.247 & 85.5 & 0.670 \\
7.044 & 233.3 & 78.6 & 2.631 & 14.0 &  0.972 \\
7.069 & 222.0 & 83.1 & 2.628 & 16.0 & 1.007 \\
7.094 & 196.0 & 115 & 2.751 & 27.2 &  1.006 \\
7.200 & 111.3 & 323 & 3.050 & 215 &  1.002 \\
\hline
\end{tabular}
}
\end{center}
\end{table}

Every computed gray vertical ring model is interpolated to a logarithmically
equidistant column density grid with 98 depth points in the range 
from $\sim 10^{-7}$\,  g cm$^{-2}$\, to\, $\Sigma_0/2$. This interpolated ring model is used as a
starting model for the calculation of a non-gray model and its emergent spectrum.

\subsection{Boundary layer vertical structure. Model atmosphere approach.}

The non-gray model is determined by Eqs.\,(\ref{u1}) and (\ref{u3}),
rewritten in the form
\be
\label{u6}
\frac{dP_{\rm g}}{dm} = g_{\rm z}-g_{\rm rad},
\ee
where $g_{\rm z}$ and $g_{\rm rad}$ are defined by Eqs.\,(\ref{u1}) and
(\ref{u8}). Here we again assume that energy is transferred by radiation alone
and this is described by the radiation transfer equation at every considered
frequency point
\be
\label{u7}
\mu \frac{dI_\nu}{d\tau_\nu} = I_{\rm \nu}-S_{\rm \nu},
\ee
where the monochromatic optical depth is determined from
\be
 d\tau_{\rm \nu} = (\sigma_e + k_\nu)\, dm,
\ee
and the source function can be expressed as
\be
  S_\nu = \frac{k_\nu}{\sigma_e+k_\nu}\, B_\nu+\frac{\sigma_e}{\sigma_e+k_\nu}\, J_\nu .
\ee
Here $\mu=\cos \theta$ is the cosine of angle $\theta$, which is the angle
between the radiation transfer direction and the ring normal. Eq.\,(\ref{u7})
determines the specific intensity $I_\nu$ at given frequency and $\mu$. The mean
intensity $J_\nu$ and Eddington flux $H_\nu=F_\nu/4\pi$ are defined by
\be \label{JH}
  J_\nu =\frac{1}{2}\, \int^{+1}_{-1} \, I_\nu\, d\mu, \qquad 
  H_\nu =\frac{1}{2}\, \int^{+1}_{-1}\mu \, I_\nu\, d\mu .
\ee

\begin{figure}
\centering
\includegraphics[angle=0,scale=1.]{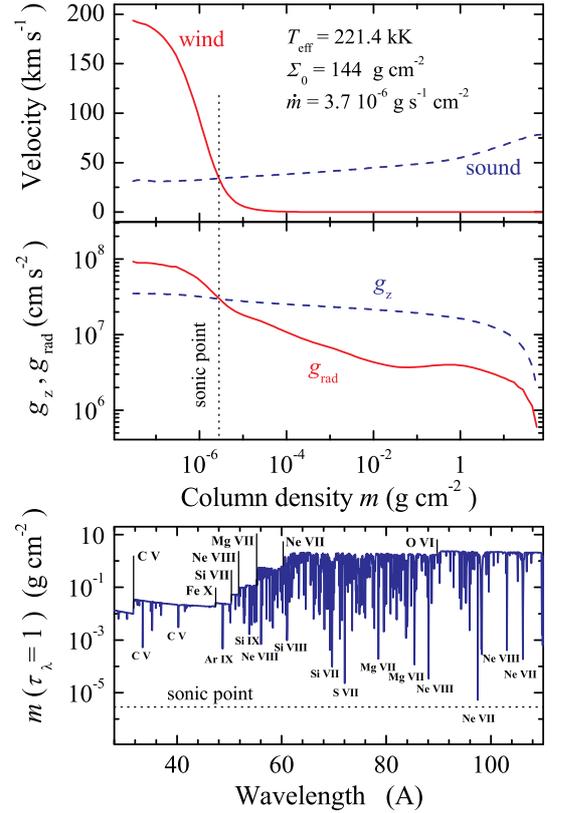}
\caption{\label{fig3}
Top panel: Sound speed (dashed curve) and wind velocity (solid curve)
distributions along depth in the ring model shown in Fig.\,\ref{fig2}. Middle
panel: Distributions of gravity (dashed curve) and radiative acceleration
(solid curve) along depth. The position of the sonic point is marked by the
dotted vertical line. Bottom panel: Depths where the emergent spectrum forms
($\tau_{\lambda}=1$). The position of the sonic point is marked by the dashed
horizontal line.}
\end{figure}

We took coherent electron scattering $\sigma_e$ into account together with the
free-free and bound-free opacities $k_{\nu}$ of all ions of the 15 most abundant
elements using opacities from \cite{VYa:95} and \citet{Verner.etal:96}. Line
blanketing was taken into account using $\sim 25\,000$ spectral lines from
the CHIANTI, Version 3.0, atomic database \citep{dere:97}.

We solve the radiation transfer equation (\ref{u7}) at three values of $|\mu|$  using 
the short characteristics method \citep{OK87}. The adopted $\mu$ values correspond to
Chebyshev-Gauss quadrature abscissas to accurately compute the integrals in (\ref{JH}).

Radiative acceleration $g_{\rm rad}$ is determined by
\be \label{u8}
g_{\rm rad} = \frac{4\pi}{c} \, 
\int^{\infty}_{0} \left[\sigma_e + k_\nu \right] H_\nu(m) \, d\nu . 
\ee
A correct model has to satisfy the energy balance equation which can be written
in two forms:
\be
 \label{u9}
4\pi \, \int^{\infty}_{0} H_\nu(m) \, d\nu = F_0\, \left(1-\frac{2m}{\Sigma_0}\right) 
\ee
and
\be
 \label{10}
4\pi\,\int^{\infty}_{0} \left[\sigma_e + k_\nu \right] \left[J_\nu-S_\nu\right] \, \, d \nu 
= -\frac{2F_0}{\Sigma_0}.
\ee
Of course, the initial model does not satisfy the energy balance equation,
having  the relative  flux error
\be
     \varepsilon_{\rm H}(m) = 1 - \frac{F_0(1-2m /\Sigma_0)}{4\pi\int_0^{\infty} H_\nu(m)\, d\nu},
\ee
and flux derivative error 
\be  \label{eq:econs1}
 \varepsilon_{\Lambda}(m) = 4\pi\,\int^{\infty}_{0} \left[\sigma_e + k_\nu \right] \left[J_\nu-S_\nu\right] \, \, d \nu 
+\frac{2F_0}{\Sigma_0}.
\ee
at each depth. It is possible to
find corrections to the temperature $\Delta T (m)$ using 
two methods modified for a non-constant bolometric flux over the ring depth
\citep{SV:92}:  the integral $\Lambda$-iteration together with a surface temperature correction 
for the optically thin parts of the atmosphere
\be
     \Delta T_{\Lambda}(m) = -  \varepsilon_{\Lambda}(m)  \ 
     \left(\int_0^{\infty}
 \left[ \frac{\Lambda_{\rm \nu, diag}-1}{1-\alpha_\nu\Lambda_{\rm \nu,diag} } \right]
k_\nu\, \frac{dB_\nu}{dT}\, d\nu  \right) ^{-1},
\ee  
where $\alpha_\nu=\sigma_e/(k_\nu+\sigma_e)$, and
$\Lambda_{\rm \nu,diag}$  is the diagonal matrix element of the $\Lambda$-operator,
and the Avrett-Krook flux correction based on the
relative flux error $\varepsilon_{\rm H}(m)$ for the optically thick parts of the ring model
\be
  \Delta T_{\rm H}(m) = - \frac{dT}{dm}\,\int_0^m\, \varepsilon_{\rm H}(x)\, dx.
\ee
Both methods were described in detail 
by \citet{Kurucz:70}.

Then
we find new values of the gas pressure using new $g_{\rm rad}$ and recalculate
the densities. Subsequently, opacities are recalculated and the radiation
transfer equations at all frequencies are resolved and a new temperature
structure is computed. Then we recalculate the geometrical depth scale $z'(m)$
using the new densities. This procedure is performed up to convergence using our
version of the computer code ATLAS \citep{Kurucz:70} that was modified to deal
with high temperatures \citep{Ibragimov.etal:03,sw:07,SMZW:13}. We assumed LTE
and accounted for pressure ionization effects using the occupation probability
formalism \citep{Hum.Mih:88}, as described by \citet{Lanz.Hub:94}.

Because of the significant effect of spectral lines on the radiation force
$g_{\rm rad}$, the atmosphere modeling approach is divided into two steps. In
the first step we compute a continuum model without spectral lines, which
provides the radiation force $g_{\rm rad} < g_{\rm z}$ at all BL ring
depths. The distributions of basic BL quantities for continuum ring models are
presented in Fig.\,\ref{fig1} by solid curves. Hydrostatic ring models cannot be
computed when spectral lines are included, because then the radiation force $g_{\rm rad}$
becomes higher than the gravity $g_{\rm z}$ in the upper, optically thin (in
continuum) layers of the models. Therefore, a line-driven wind has to arise from
the BL surface, which is similar to the line-driven winds from hot stars
\citep{CAK:75, KudrP:00}. This kind of wind also exists for CV accretion disks
\citep{DrewV:85, PriRos:95, Proga98}.

\subsubsection{Line-driven wind approach}

Here, we are not interested in the properties of the wind itself, because it
probably effects only the strongest spectral lines but not the continuum or
moderate and weak spectral lines. In previous works \citep{Ibragimov.etal:03,
SMZW:13}, where hot WD model atmospheres were computed, the gas pressure was
artificially fixed to 10\% of the total pressure at the upper layers, where
$g_{\rm rad} > g_{\rm z}$. The condition $P_{\rm gas} = 0.1P$ was chosen because
it was approximately correct at those atmospheric layers, where $g_{\rm rad}
\approx g_{\rm z}$. This condition physically means that we assume some
artificial wind velocity law that satisfies the imposed
condition. Unfortunately, this simple approach is not working for the BL case
because gravity $g_{\rm z}$ depends on the vertical coordinate $z$ (it is constant
for ordinary stellar atmospheres). Therefore, we develop another simple
hydrodynamic approach which takes into consideration an expansion of the upper
ring layers. This approach is not self-consistent as we do not include effects
of motion in the radiation transfer. Thus we consider a hydrodynamical model for
the ring upper layers but the radiation pressure force governing this expansion
is computed for a formally static medium. This approach allows computing the
model emergent spectra but the obtained wind properties such as the local
mass-loss rate and wind-velocity distributions are not completely correct and
cannot be considered as reliable results.

The considered approach is based on Euler's equation 
\be\label{u11}
\frac{d}{dz}(\dot m {\rm v} + P_{\rm g}) = -\rho (g_{\rm z} - g_{\rm rad}),
\ee 
where $\dot m$ is the local mass-loss rate ([$\dot m$]= g s$^{-1}$ cm$^{-2}$)
and ${\rm v}$ is the gas velocity at given $z$. This equation can be rewritten as
\be \label{u12}
\dot m \left(1-\frac{a^2}{{\rm v}^2}\right)\frac{d{\rm v}}{dz} = 
-\rho (g_{\rm z} - g_{\rm rad}) - \rho \frac{da^2}{dz},
\ee
where $a$ is the sound speed connected with gas pressure and matter density by
\be
 P_{\rm g}=\rho a^2. 
\ee 
Here we used the continuity equation for the plane 1D motion
\be \label{u12a}
 \dot m = \rho {\rm v}.  
\ee
It is well known \citep[see, e.g.][]{Mih:78} that a correct solution of Eq.\,(\ref{u12}) 
has to pass through a singular point with $|{\rm v}| =
a$, where the left and the right side of the equation
vanish simultaneously. Therefore, these two conditions must be fulfilled in the
singular point:
\be \label{u13a}
a^2={\rm v^2},
\ee
and
\be \label{u13}
 \frac{da^2}{dz} = -(g_{\rm z} - g_{\rm rad}),
\ee
or
\be \label{u13a}
 C = \frac{da^2}{dz} + (g_{\rm z} - g_{\rm rad})=0.
\ee
We suggest the following scheme to compute a model of the BL ring with a
line-driven wind. We take the model computed without lines as an initial model
and solve the radiation-transfer equation with spectral lines taken into
account. As a result we obtain that the radiative acceleration $g_{\rm rad}$ is
greater than the current gravity at the upper layers. At this stage we have the
gas pressure $P_{\rm g}(m)$ and the gas density $\rho(m)$ distributions from the
continuum model, and the radiative acceleration $g_{\rm rad}(m)$ and the gas
temperature $T(m)$ distributions after the first temperature-correction
iteration. The geometrical depth scale $z(m)$ and corresponding gravity $g_{\rm
z}$ are connected with the gas density of the initial (continuum) model. Using
these distributions we can find the sound-speed distribution $a(m)$ and its
derivative $da(m)/dz$, and then find the depth $m_{\rm c}$ where
Eq.\,(\ref{u13}) is satisfied. We know at this depth the wind velocity must be
equal to the sound speed $a(m_{\rm c})$. Therefore, we can evaluate the local
mass-loss rate
 \be \label{mt}
 \dot m = \rho(m_{\rm c})\, a(m_{\rm c})/2
 \ee 
using the continuity equation (\ref{u12a}). Here we have taken into account that
the correct gas density $\rho_{\rm w}$ in the singular point for the model with
the wind must be half of that in the hydrostatic model $\rho_{\rm st}$, see
Eq.\,(\ref{u11})
\be
 \dot m a +\rho_{\rm w} a^2 = \rho_{\rm w} a^2 + \rho_{\rm w} a^2 = \rho_{\rm st} a^2.
\ee 
Using this mass-loss rate $\dot m$ we solve two ordinary differential equations 
\be \label{u15}
\frac{d\,P_{\rm w}}{d\,z} = -\rho(g_{\rm z} - g_{\rm rad})
\ee
and
\be \label{u16}
 \frac{dm'}{dz}=\rho,
\ee
where $P_{\rm w} = \dot m {\rm v} + \rho a^2$, on a new fine and equidistant
geometrical depth grid $z$ (50\,000 points) from the midplane ($z=0$) up to the
surface ($m'=\Sigma_0/2$) with the boundary conditions ${\rm v}=0$, $m'=0$, and
$P_{\rm w}(0)$, $\rho(0)=P_{\rm w}(0)/a(0)^2$ at the midplane, using the
shooting method. The current gas density is calculated using the
current $P_{\rm w}$
\be \label{u17}
 \rho = \frac{P_{\rm w}}{2a^2} \pm \sqrt{\left(\frac{P_{\rm w}}{2a^2}\right)^2-
 \left(\frac{\dot m}{a}\right)^2}. 
\ee
The plus sign is used for wind velocities below the sonic point and the minus
sign in the supersonic part of the model. The wind velocity is found using the
continuity equation (\ref{u12a}). The necessary current values of the $g_{\rm
rad}(m')$ and $da^2(m')/dz = \rho d\,a^2(m')/d\,m'$ are found using a spline
interpolation in the dependencies $g_{\rm rad}(m)$ and $da^2(m)/dm$ known from
the previous iteration, taking $m=\Sigma_0/2-m'$.

Usually, the necessary conditions Eqs.\,(\ref{u12}) and (\ref{u13}) are not
satisfied at the sonic point for the first tried initial condition for $P_{\rm
w}(0)$. Therefore, we find the necessary $P_{\rm w}(0)$ with the dichotomy
method. Usually, the correct initial value of the gas pressure at the midplane
differs by a few percent from the gas pressure at the midplane found in the
previous iteration. We adopt the two following numerical convergence criteria at
the singular point
$$
 \frac{a-|{\rm v}|}{a}<10^{-2}, ~~~~~~~~~~~~ \frac{C}{g_{\rm z}}< 10^{-2}.
$$ 
After obtaining the solution, all quantities are interpolated on the model
atmosphere column density grid $m$ (98 points) which is used for the radiation
transfer solution. Therefore, we have a new distribution of the gas pressure
$P_{\rm g}=\rho a^2$ with the line-driven wind taken into account which is in
accordance with the radiative acceleration $g_{\rm rad}$. Then we find a new
gas density distribution using the new $P_{\rm g}$ and the temperature
distribution obtained after the temperature-correction iteration. For this aim
the ideal gas equation (\ref{igl})
is used together with charge and number density conservation laws. Then the
radiation transfer is re-solved and a new temperature correction is
performed. The above described procedure to find a new gas-pressure distribution
is repeated with a new radiative-acceleration distribution. There is only one
difference. The local mass-loss rate is now calculated (see Eq.\,\ref{mt}) without
division by 2 because the current model takes into consideration the line-driven
wind.

This iteration scheme is converging and it is stable. As a result we get a
self-consistent model of the BL ring and its emergent spectrum. Results of
calculations for the reference ring model
are presented in Figs.\,\ref{fig2} and \ref{fig3}. The emergent spectrum
(Fig.\,\ref{fig2}, top panel) is similar to hot stellar atmospheres with similar
effective temperatures \citep[see e.g.][] {SMZW:13} and shows the same
absorption edges as the continuum spectrum together with a forest of spectral
absorption lines. The temperature structure (Fig.\,\ref{fig2}, bottom panel)
differs significantly from the gray temperature distribution mainly because of
line-blanketing effects. The final model's wind velocity law (Fig.\,\ref{fig3},
top panel) and the radiation force $g_{\rm rad}$ distribution (Fig.\,\ref{fig3},
middle panel) are shown together with the sound speed $a$ and gravity
$g_{\rm z}$ distributions. The sonic point is seen to be located at the upper
ring layers ($m \approx 10^{-5}$~g cm$^{-2}$) and the corresponding mass-loss
rate is sufficiently low ($\dot m \approx 10^{-5} - 10^{-6}$~g
s$^{-1}$~cm$^{-2}$). Formally, the sonic point is above the formation depths of
even the strongest spectral lines (Fig.\,\ref{fig3}, bottom panel) and,
therefore, the wind does not affect the spectrum.
 
\begin{figure}
\centering
\includegraphics[angle=0,scale=1.]{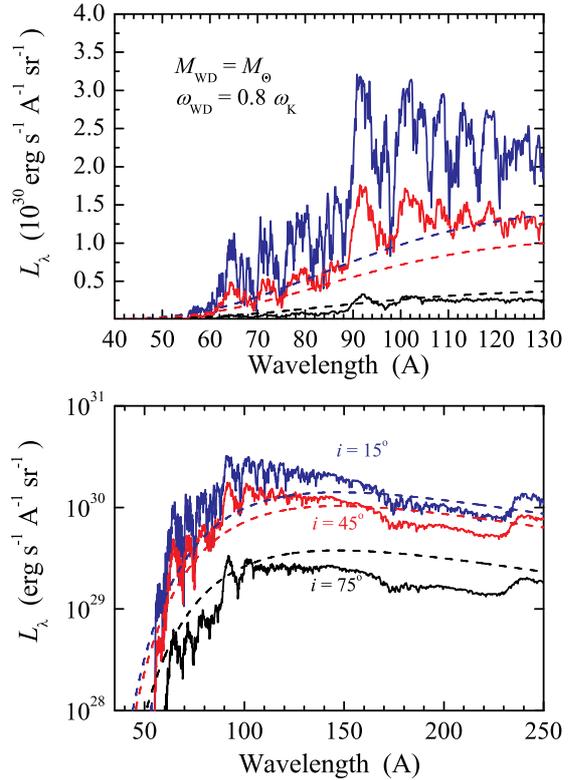}
\caption{\label{fig4}
Spectra of the BL model $M = M_\odot$, $\omega_{\rm WD} = 0.8\,\omega_{\rm K}$ for 
three inclination angles relative to the line of sight. The corresponding blackbody 
approximations are shown by dashed curves.
 }
\end{figure}

\begin{figure}
\centering
\includegraphics[angle=0,scale=1.]{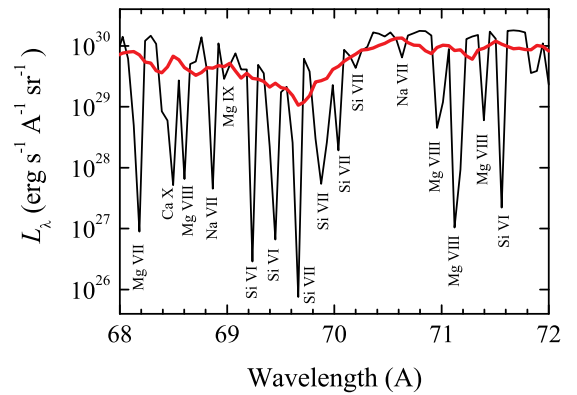}
\caption{\label{fig4a}
{Detail of the integral emergent fluxes of the same BL model as in Fig.\,\ref{fig4} 
computed for $i$ =0$\degr$ (rotation does not affect the spectrum, thin black curve),
and 15$\degr$ (thick red curve). }
}
\end{figure}

\begin{figure}
\centering
\includegraphics[angle=0,scale=1.]{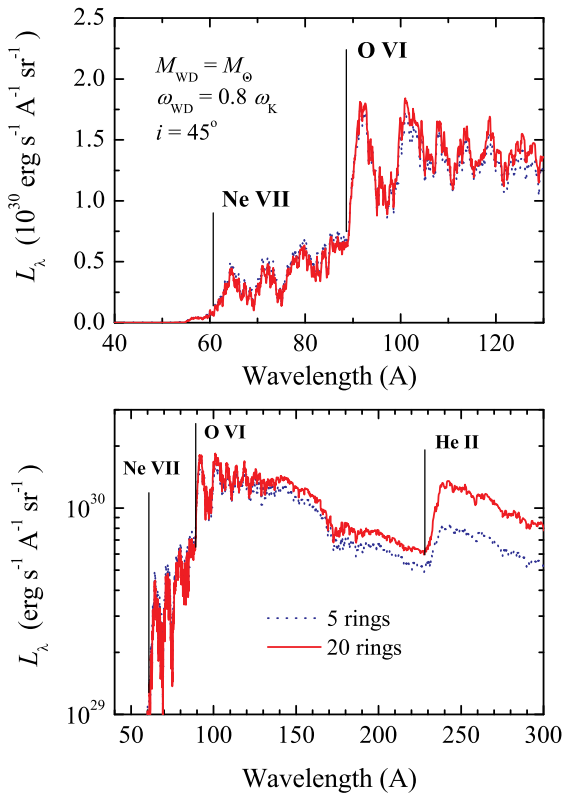}
\caption{\label{fig5}
Comparison of the spectra of the same BL model (see Fig.\,\ref{fig4}) computed using five (dotted curve)
 and twenty (solid curve) rings.
}
\end{figure}

\begin{figure}
\centering
\includegraphics[angle=0,scale=1.]{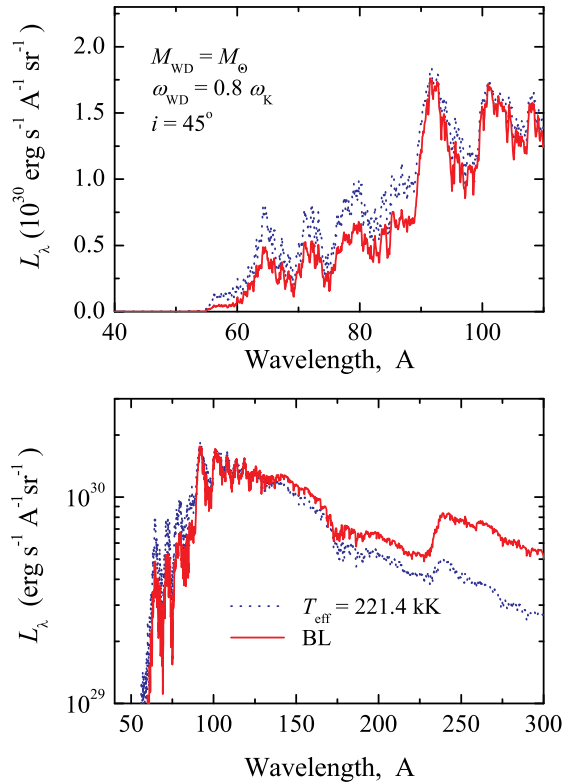}
\caption{\label{fig6}
Comparison of the BL model spectrum from Fig.\,\ref{fig4} (solid curve) with the 
spectrum of the hottest ring (dotted curve) normalized to the flux at 100~\AA. 
}
\end{figure}
 
\subsection{Integral BL spectra}

When the computation of all ring model spectra is finished, we integrate them over the BL
radial coordinate with taken Doppler broadening due to ring rotation  into
account
\bea \label{int_sp}
 L_{\lambda} &=& \cos i \int_{\rm R_{\rm in}}^{R_{\rm out}} \, R\, dR\, \int_0^\pi \, I_{\lambda'}(\cos i)\, 
 d\varphi \nonumber \\ &\approx&
  \cos i\, \sum_{\rm j=1}^{\rm N_{\rm R}} \,R_{\rm j}\, \Delta R_{\rm j} 
 \sum_{\rm k=1}^{\rm N_{\varphi}}\, \Delta \varphi_{\rm k}\, I_{\lambda'}(\cos i),
\eea
where $i$ is the inclination angle of the BL relative to the line of sight,
$I_{\lambda}(\cos i)$ is the local specific intensity in the direction of the
line of sight, $N_{\rm R}$ is the number of considered BL rings (5 or 20), and
$N_{\varphi} = 100$ is a number of considered ring sectors. Here $\lambda'$ is the
Doppler shifted wavelength:
\be
\lambda' = \lambda + \frac{{\rm v}_{\varphi}(R)}{c} \sin i\, \cos\varphi.
\ee 
The azimuthal velocity distribution ${\rm v}_{\varphi}(R)$ is taken from the
1D BL model. We note that integration is only performed over the visible
BL part.  We assume that we see only a quarter of the total BL surface,
which is situated in front of the WD on the visible part of the disk, and that the visible part does not
depend on the inclination angle. The change of the projection area of the BL's visible part is proportional to
$\cos i$.
Specific intensities for every ring model are  computed using Eq. (\ref{u7}) for six
angles relative to the normal: 0, 15, 30, 45, 60 and 75 degrees. Therefore,
integral spectra can be computed for these inclination angles relative to the
line of sight. We ignored any relativistic effects because they are low.

\section{Results}

Using the method presented above, we computed spectra of several BL models.
Examples of the integral reference BL model spectra for three angles are shown
in Fig.\,\ref{fig4}. These spectra are calculated using the BL model
partitioned in 20 rings. The spectra show broad absorption and quasi-emission
features. The absorption features arise due to rotational broadening and
blending as a result of many absorption lines at this place of the spectrum. The
quasi-emission features are continuum sections with a low amount of absorption
lines. These statements are illustrated by Fig.\,\ref{fig4a}, where a narrow
section of the integral spectra computed for $i$ = 0$\degr$ and 15$\degr$ are
shown. In Fig.\,\ref{fig4} the spectra computed in blackbody approximation are
also shown. We focus attention on the fact that the atmosphere-model
intensities compared to the blackbody spectra are different for low and high
inclination angles. This is the result of different limb-darkening laws for the
ring atmospheres (the intensity is mainly concentrated along the normal)
compared to the isotropic blackbody radiation.

We investigated the importance of the number of rings to
describe the model spectra. In addition to the spectrum of the BL reference model 
presented above, we computed the spectrum of the same model using five rings. 
A comparison of both spectra is shown in Fig.\,\ref{fig5}. The
difference is significant at wavelengths longer than 120 -- 130~\AA, but the
spectra are almost identical at the shorter wavelengths. Our work was
particularly motivated by the {\sl Chandra} X-ray spectrum of SS Cyg in outburst
\citep{Mauche04}. The grating spectra cover the wavelength range 10 -- 130~\AA,
therefore, to model the observed BL spectra five rings are sufficient.
Moreover, even the hottest ring spectrum alone can approximate the integral BL
spectrum with a relatively good accuracy (see Fig.\,\ref{fig6}). All BL spectra
of the two grids described above were computed using five rings.

\begin{figure}
\centering
\includegraphics[angle=0,scale=1.]{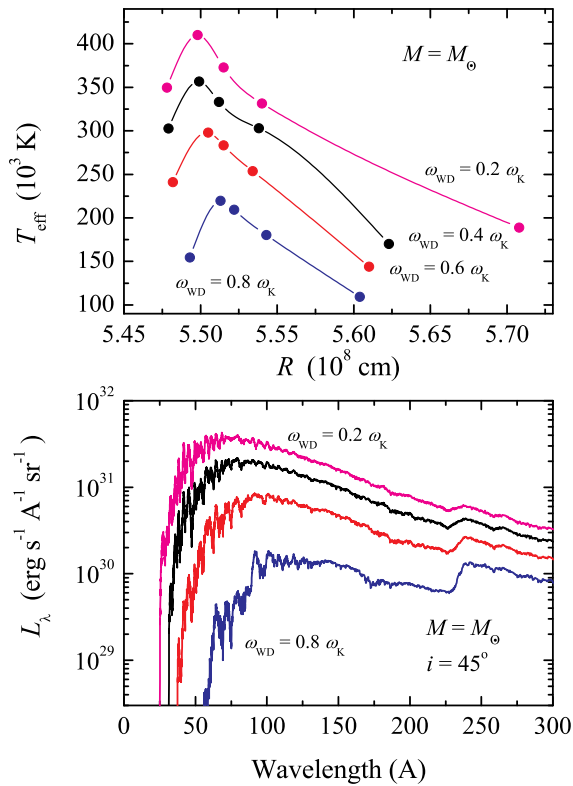}
\caption{\label{fig7}
Effective temperature distributions (top panel, the temperatures of the
considered rings are marked by dots) and model spectra (bottom panel) for
the models with different angular velocities (0.2, 0.4, 0.6 and
0.8\,$\omega_{\rm K}$) and  fixed WD mass ($M_{\rm WD} = M_\odot$).
}
\end{figure}

\begin{figure}
\centering
\includegraphics[angle=0,scale=1.]{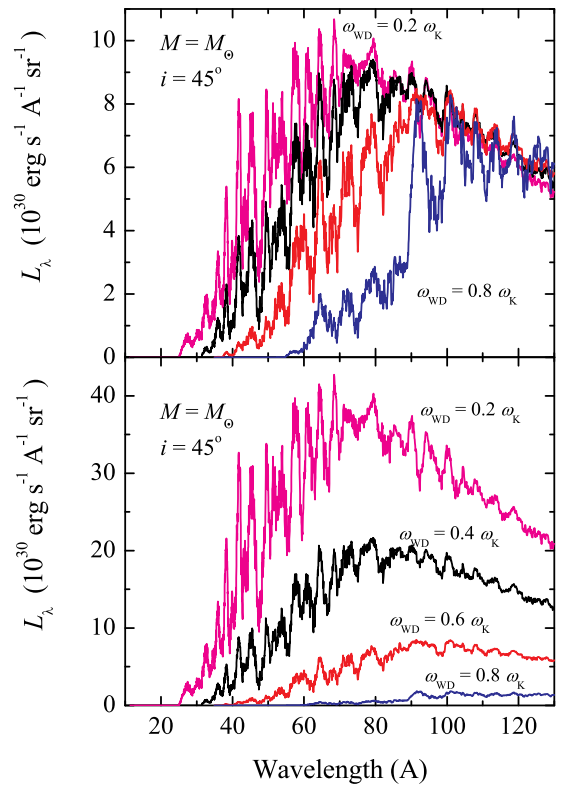}
\caption{\label{fig8}
Spectra of the same models as in Fig.\,\ref{fig7} in the observed wavelength
range. The spectra in the top panel are normalized to the $\omega_{\rm WD}$ =
0.6\,$\omega_{\rm K}$ model spectrum at 100~\AA.
}
\end{figure}

\begin{figure}
\centering
\includegraphics[angle=0,scale=1.]{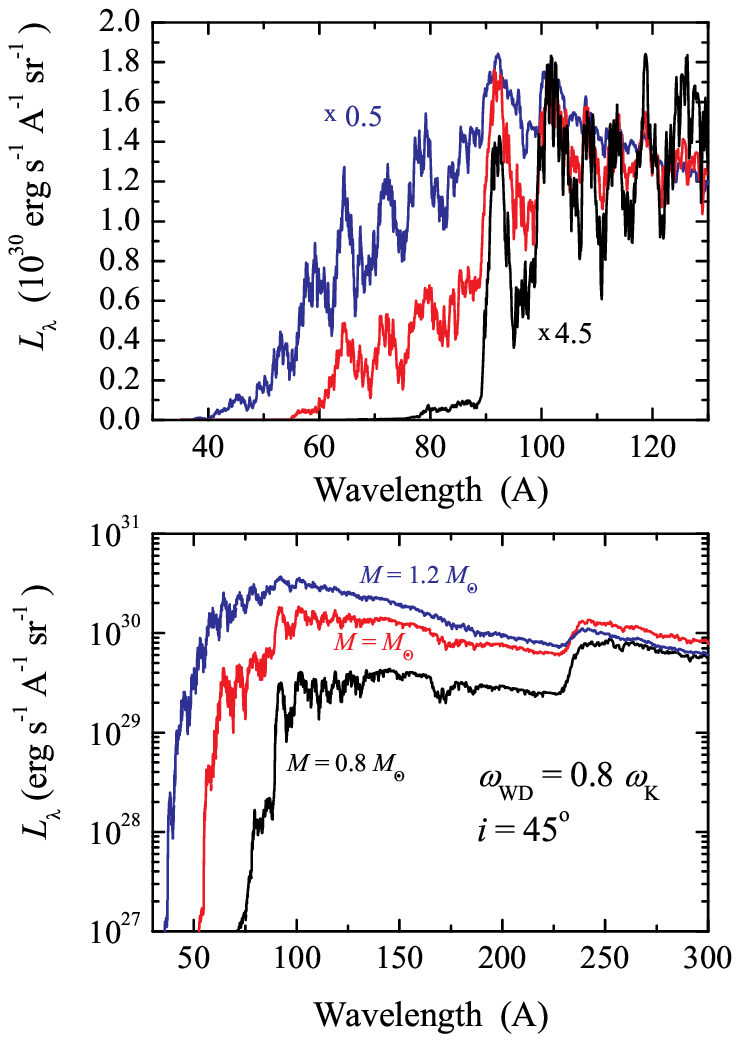}
\caption{\label{fig9}
Comparison of BL model spectra for three WD masses (0.8, 1, 1.2 $M_\odot$) and
fixed relative angular velocity ($\omega_{\rm WD} = 0.8\,\omega_{\rm K}$). The
spectra in the top panel are normalized to the $M_{\rm WD} = M_\odot$ model
flux at 100~\AA.
}
\end{figure}

It is important to explore the dependence of the BL model spectra on the BL parameters.
For this purpose, we computed spectra of two model grids. In the first one we fixed
the relative WD angular velocity $\omega_{\rm WD} = 0.8\,\omega_{\rm K}$ and
investigated how the spectra depend on the WD mass $M_{\rm WD}$. In the second
set we fixed the WD mass $M_{\rm WD} = M_{\odot}$\, and varied the relative WD
angular velocity. The obtained results are described below.

A comparison of the computed integral BL spectra for various parameters is shown
in Figs.\,\ref{fig7} -- \ref{fig9}. The basic differences of the presented
spectra are determined by differences in the BL model effective temperatures
(see e.g. Fig.\,\ref{fig7}, top panel). The radial widths of all the computed
models are similar \citep[see the same Figure and ][]{HKSW:13}, but the
luminosities can change more significantly. So, the BL effective temperatures
directly connect with the BL luminosities, which are defined by mass accretion
rate $\dot M$, WD mass, and relative WD angular velocity $\omega_{\rm
WD}/\omega_{\rm K}$ \citep{Kluzn87,Kley91}:
\be 
\label{bl_lum}
 L_{\rm BL} = \frac{1}{2}\frac{GM_{\rm WD}\dot M}{R_{\rm WD}}
 \left(1-\frac{\omega_{\rm WD}}{\omega_{\rm K}}\right)^2.
\ee 
All the models considered here have a fixed mass accretion rate. Therefore,
luminosities and effective temperatures of these models only depend on WD masses and
relative angular velocities. Both of them increase with the WD mass
and decrease with the relative angular velocity. Higher BL effective
temperatures have to correspond to harder spectra and higher ionization degree
of the matter.

Our computations confirm these qualitative conclusions. Emergent spectra of the
BL models around a solar mass WD with various relative angular velocities are
shown in Fig.\,\ref{fig7}. The effective temperatures increase with decreasing WD
relative angular velocity, and the emergent spectra become harder.
As a result, the photo-absorption edges and the intensities of quasi-absorption
and quasi-emission features are changing. In particular, the position of a
notable spectral short-wavelength roll-off shifts from 60~\AA\ in the spectrum
of BL model around the fastest rotating WD up to 25~\AA\, in the spectrum of the
BL model around the slowest rotating WD. The same spectra in the 15 --
130~\AA\, wavelength band on a linear flux scale are shown in Fig.\,\ref{fig8}.
The spectra, which are normalized at 100~\AA, are also shown in the top panel of
the same Figure to emphasize the changes in the spectral shapes.

We note that some ring models 
(near the minimum of the surface density distribution) of the BL model around the non-rotating WD 
($M_{\rm WD} = M_{\odot}$, $\omega_{\rm WD} = 0$) did not
converge and a corresponding integral BL model spectrum was not computed. The
reason is that these rings are effectively optically thin 
($\tau_{\rm eff, \nu} = \int_0^{\Sigma_0/2}\sqrt{\sigma_{\rm e}(\sigma_{\rm e}+k_\nu)}\,dm\, < 1$) 
over a wide wavelength
range. We can see that the average Rosseland opacities for the hottest rings for
the BL models  around slowly rotating WDs are close to pure electron scattering
and the input of the ''true'' opacity $k_{\rm kr}$ is negligible, because $\tau_0 \approx \sigma_{\rm e} \Sigma_0/2$ 
(see Table \ref{tabl1}).  Possibly, the similar effect may exist for other BL models around non-rotating
WDs. It could be less important for the BL around WDs with lower masses and could be
reduced for lesser mass accretion rates.
 
The difference between BL spectra computed for the fixed WD relative angular
velocity but various WD masses are qualitatively similar (Fig.\,\ref{fig9}).
The spectrum of the BL model around the heaviest WD is harder than the spectra
of BL models around less heavy WDs.

\section{Comparison with observations}

As mentioned in the introduction, there are only four dwarf novae with observed
soft X-ray/EUV radiation in outbursts, which can be associated with optically
thick BLs. One of them is SS~Cyg \citep{Mauche04,SMZW:13}. Fitting the spectrum
gave effective temperature estimations  of 190 (250)~kK and a bolometric BL
luminosity 18 (5)~$\times 10^{33}$\, erg s$^{-1}$ for an assumed distance of
160~pc. The bolometric disk luminosity of SS Cyg at the peak of the outburst,
$L_{\rm D}\approx 10^{35}$\, erg s$^{-1}$, 
was estimated by \citet{Mauche04} using the same distance and the data published by \citet{PH:84}.
We note that the soft X-ray spectrum of SS Cyg was obtained by {\sl Chandra} at the
outburst peak, too. The corresponding mass accretion rate $\dot M\approx 1.5 \times 10^{-8}\, M_\odot$/yr,
which was used for the computation of all the BL models in this work, was obtained using this luminosity
for the adopted WD parameters, $M_{\rm WD} = M_{\odot}$ and $R_{\rm WD} = 5.5\times10^8$~cm. 
The corresponding ratio of BL luminosity $L_{\rm BL}$ to disk luminosity
$L_{\rm D}$ is 0.18 (0.05). These values 
correspond to the fit of the spectrum by a hot stellar atmosphere that is close
to Eddington limit \citep[with $\log g =6.2$, ][]{SMZW:13}, and the 
values in brackets were obtained using blackbody fits with broad absorption lines
\citep{Mauche04}.
The recent determination of SS Cyg's distance by \citet{NB:13} gave a lesser value, 
$\sim$ 110 pc instead of $\sim$160 pc.
It reduces the disk and the BL luminosities mentioned before, but does not change their ratio.  The low value 
of $L_{\rm BL}/L_{\rm D}$ supports the hypothesis
about a fast WD rotation in SS~Cyg \citep{Mauche04}. In the ultraviolet spectrum
of this CV in quiescence the absorption lines, which are in principle available
to determine the WD's projected rotation velocity, were not found
\citep{Sion.etal:10}.  The most probable reason for this finding are strong emission lines in the
 UV spectrum of SS Cyg in quiescence \citep{Long.etal:05}, which could mask absorption 
 lines in the WD spectrum.
 But it is necessary
to keep in mind that the BL luminosity estimations are very model dependent and
have large uncertainties.

\begin{figure}
\centering
\includegraphics[angle=0,scale=1.]{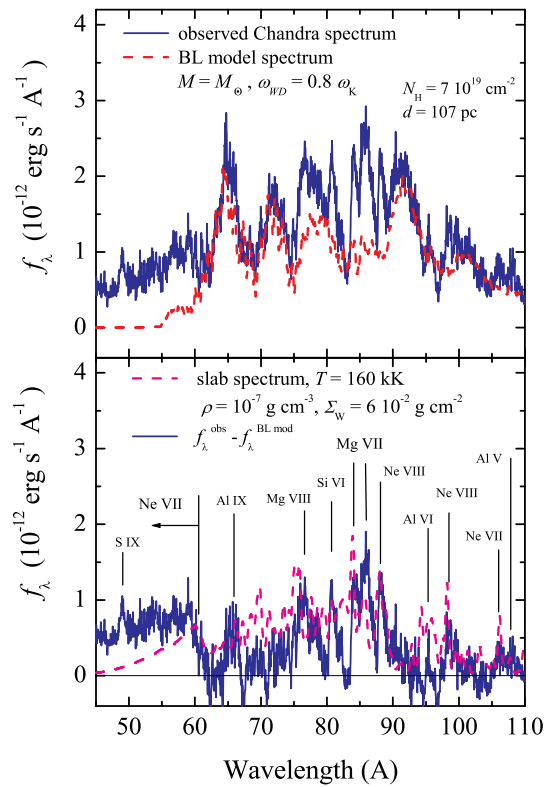}
\caption{\label{fig11}
Top panel: Comparison of the observed SS~Cyg soft X-ray spectrum (solid curve) with the BL model spectrum for a 1~M$_{\odot}$ WD
and relative angular velocity $\omega_{\rm WD} = 0.8\,\omega_{\rm K}$ (dashed curve).
Bottom panel: Subtracted (observed minus BL model) spectrum (solid curve) together with the homogeneous slab model
spectrum (dashed curve). The emission continuum of Ne VII below 60 \AA\,  and the strongest identified emission lines are marked.
}
\end{figure}

 \begin{table}[htbp]
\begin{center}
\caption{Parameters of the strongest emission lines, preliminarily  identified in the subtracted spectrum. Part of them could be blends with other lines. 
Data were taken from the {\sc CHIANTI} database \citep{dere:97}.
}\label{lines}
{
\begin{tabular}{llll}
$\lambda$, $\AA$ & Ion &  Transition &$gf$      \\ 
\hline
49.12 & S IX  & 2p$^4$ $^3$P$_2$ - 3d $^3$D$_3^{\rm o}$ &1.12 \\
65.7 & Al IX &  2p$^3$ $^2$P$_{3/2}$ - 3d $^2$D$_{5/2}$ & 1.33 \\
76.77 & Mg VIII & 2p$^3$ $^4$S$^{\rm o}$ - 3d $^4$P & 4.7$^a$ \\
80.5 & Si VI & 2p$^5$ $^2$P$^{\rm o}$ - 3d $^2$P  &2.3$^a$ \\
83.97 & Mg VII   &2p$^2$ $^3$P - 3d $^3$D$^{\rm o}$ &$\sim$ 7$^b$ \\
85.41 & Mg VII   & 2p$^2$ $^1$D - 3d $^1$F$^{\rm o}$&4.23 \\
88.1 & Ne VIII  & 2s $^2$S - 3p $^2$P$^{\rm o}$&0.6$^a$ \\
98.26 & Ne VIII  & 2p $^2$P$_{3/2}^{\rm o}$ - 3d $^2$D$_{5/2}$&2.28 \\
106.14 & Ne VII   &2p $^3$P$^{\rm o}$ - 3d $^3$D  &4.3$^a$ \\
107.94 & Al V   &2p$^5$ $^2$P$_{3/2}^{\rm o}$ - 3d $^2$D$_{5/2}$  & 1.4 \\
\hline
\end{tabular}
}
\end{center}
\begin{center}
$^a$ - doublet, the $gf$ values are summed. \\
$^b$ - triplet, the $gf$ values are summed. 
\end{center}
\end{table}

\begin{figure}
\centering
\includegraphics[angle=0,scale=1.]{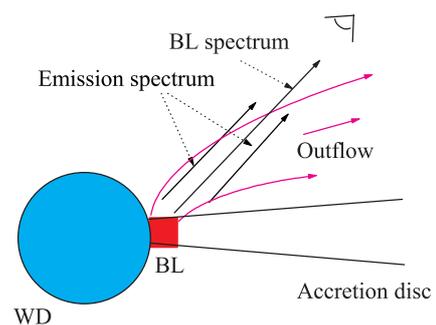}
\caption{\label{fig13}
 Scheme of the BL outflow. 
}
\end{figure}

\begin{figure}
\centering
\includegraphics[angle=0,scale=1.]{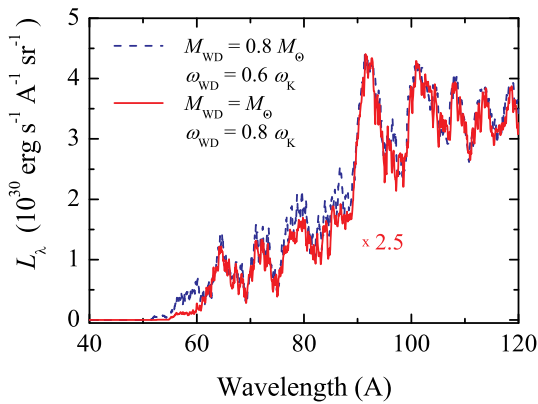}
\caption{\label{fig12}
Comparison of the BL model spectra for a 0.8~M$_{\odot}$ WD
and relative angular velocity $\omega_{\rm WD} = 0.6\,\omega_{\rm K}$
(dashed curve)
and  for a 1~M$_{\odot}$ WD
and relative angular velocity $\omega_{\rm WD} = 0.8\,\omega_{\rm K}$
(colid curve). The flux of the latter spectrum is multiplied by 2.5. 
}
\end{figure}
 
We compare our BL model spectra to the observed {\sl Chandra} spectrum of SS Cyg \citep{Mauche04}, using the well known
normalization for converting the calculated spectra to the spectrum at Earth 
\be
f_{\lambda} = \frac{L_{\lambda}}{d^2},
\ee
where $d$ is the distance. A relatively good fit was obtained for 
 the model spectrum of a BL with parameters $M_{\rm
WD} = M_{\odot}, \omega_{\rm WD} = 0.8\, \, \omega_{\rm K}$ for the distance $d \approx$\, 107 pc and the absorption 
by the neutral interstellar medium  with $N_{\rm H} \approx 7\times10^{19}$\,  cm$^{-2}$  (Fig.\,\ref{fig11}).
We see that the model spectrum describes some prominent features in the observed spectrum, but there are also some 
additional features. The subtracted (observed minus BL model) spectrum (Fig.\,\ref{fig11}, bottom panel) is
very similar to  a spectrum of an optically thin hot plasma with a strong emission continuum of  NeVII and numerous broad
emission lines. Part of them are identified (see Fig.\,\ref{fig11} and Table\,\ref{lines}) as spectral lines with great $gf$ values.
It is most probable that the difference spectrum is due to an optically thin outflow, which is projected on the cool WD
and the accretion disc, see Fig.\,\ref{fig13}.  We also show in the bottom of Fig.\,\ref{fig11} the absorbed (with the same $N_{\rm H}$) 
spectrum of the homogeneous slab computed in LTE approximation
with temperature $T = 160$\, kK,  density $\rho = 10^{-7}$\, g cm$^{-3}$, and 
 surface density $\Sigma_{\rm W} = 6\times10^{-2}$\, g cm$^{-2}$.
The corresponding geometrical thickness of the slab is $6\times10^5$\, cm. 
The spectrum is computed by the short characteristics method and broadened  using a Gauss function with
$\sigma = 0.2 \,\AA$. This model slab spectrum cannot fit all the emission details
correctly, because the real outflow has to be strongly inhomogeneous, and, probably, the LTE approximation is not correct there.
But it reproduces the general shape of the subtracted spectrum supporting our hypothesis. We found that the emission area of
this slab is approximately four times greater than the visible model BL area. If we assume that the obtained  density 
of the slab is correct for the averaged density of the outflow, we can estimate a total mass loss rate of
$\dot M_{\rm W} \sim \rho {\rm v}_{\rm esc} S_{\rm BL} \sim 10^{-7}$\, g cm$^{-3}$ \,
$\times\, \sim 10^{8}$\, cm s$^{-1}$\, $ \times \sim10^{16}-10^{17}$ cm$^{2}$\, $\sim 10^{17}-10^{18}$\, g s$^{-1}$. 
This guess is very crude, but, nevertheless,  it is comparable with the mass accretion rate, 
and, therefore, future BL models have to be computed with
 mass loss taken into account.

 However, we have to point out that a recent estimation gave a lesser WD mass in SS Cyg, 
$M_{\rm WD} = 0.81 \pm 0.18\, M_\odot$ \citep{BRB:07} instead of 1\,$M_\odot$.
We may expect that some BL model around a slower rotating WD with $M_{\rm WD} = 0.8 M_\odot$
can have similar effective temperatures (and a similar spectrum) 
as the considered BL model with  $M_{\rm WD} =  M_\odot$ and
$\omega_{\rm WD} = 0.8 \omega_{\rm K}$. Indeed, we find that the spectra of BL models with
parameters  $M_{\rm WD} = 0.8 M_{\odot}, \omega_{\rm WD} = 0.6\, \omega_{\rm K}$
and $M_{\rm WD} = M_{\odot}, \omega_{\rm WD} = 0.8\, \omega_{\rm K}$ are very similar
(see Fig.\,\ref{fig12}). 
Therefore the spectrum of the BL model with
parameters $M_{\rm WD} = 0.8\, M_{\odot}, \omega_{\rm WD} \approx 0.6\, \omega_{\rm
K}$ also has to fit the observed SS Cyg soft X-ray spectrum in the outburst with $d \approx 170$\, pc. 
The ratios $L_{\rm BL}/L_{\rm D}$ are 0.04 
and 0.16 for these BL models, respectively, and they are close to the estimations obtained by \citet{Mauche04} 
and \citet{SMZW:13}. Therefore, the BL
model spectra computed here can fit the soft X-ray spectrum of SS~Cyg in
outburst. 

The second bright dwarf nova that exhibited a prominent soft X-ray/EUV flux
during outburst is U~Gem \citep{CM:84, Longetal:96}. But the properties of this
emission differ significantly from the case of SS~Cyg. A blackbody fit gives a
temperature about 140~kK and a BL luminosity of about 4$\times
10^{34}$\,erg\,s$^{-1}$ \citep[for the distance 90~pc, ][]{Marsh.etal:90} with
the corresponding ratio $L_{\rm BL}/L_{\rm D} \sim 0.5$ \citep{Longetal:96}.  
The earlier X-ray observations performed by  the {\sl Einstein} observatory gave
 similar relations between the blackbody temperatures and the observed fluxes 
of these CVs' soft X-ray spectra \citep{Cordova.etal:80b, Cordova.etal:80a, CM:84}. 

The estimated
rotation velocity of the WD in this CV is very low \citep[$<
100$\,km\,s$^{-1}$, ][]{Sion.etal:94} and the derived relative BL luminosity is
in accordance with the expected relative BL luminosity (see
Eq.\,\ref{bl_lum}). But all 1D BL models including ours predict higher
effective temperatures for this luminous BL, at least 200 -- 300~kK, in
contradiction with the observed temperature. This means that the BL emitting
area in U~Gem is about 5 -- 10 times greater than in SS~Cyg, i.e. the BL width
can be tens of percent of the WD radius instead of a few percent. Therefore, a
new BL model is necessary to explain this fact. 

A correct theory of optically
thick BLs has to explain the observational properties of both dwarf novae.
 A possible reason could be the 2D nature of BLs around slowly rotating WDs.
For example, the BL matter can spread over the WD surface increasing the radiating area
\citep{Kley91, FB:05, Balsara.etal:09}. Another possible physical mechanism is deposition
of a significant part of BL energy deep into the WD \citep{KT:78, GRS:95}, as it was also considered for T Tau type stars
 \citep{BR:92, RB:95}, with subsequent radiation of this
energy by a significant part of the WD surface above and below the equatorial plane.
 An additional possibility was suggested by \citet{IS:96}, who proposed that a significant part
of the energy released in the BL might be carried away by some outflow (wind).

The EUV spectrum of U~Gem in outburst is very similar to the soft X-ray spectrum
of SS~Cyg in outburst. It also shows numerous broad absorption- and
emission-like features, but a few strong emission lines are also detected
\citep{Longetal:96}. The latter probably arise in a strong BL wind and they are
visible due to a higher inclination of U~Gem's orbital plane relative to the
line of sight \citep[$i \approx 70\degr$,][]{ZhR:87} in comparison with the
inclination angle of SS~Cyg \citep[$i \approx 50\degr$,][]{BRB:07}. The even
more inclined SU~UMa-type dwarf nova OY~Car \citep[$i \approx
83\degr$,][]{Lit.etal:08} shows an EUV spectrum dominated by prominent emission
lines in super-outburst as presented by \citet{MR:00}. These authors evaluated
the wind mass-loss rate ($\le 10^{-10} M_{\odot}$\,yr$^{-1}$) and argued that
this value cannot be explained by a line-driven disk wind. It is possible that
a line-driven BL wind as presented in our calculations can help to resolve this
problem.  We suggest that radiation-driven winds are much more powerful for BLs around slowly
rotating WDs because of much higher effective temperatures at the same emitting area (see, e.g., 
Table \ref{tabl2}). Therefore, such winds could be optically thick like the winds of Wolf-Rayet stars
\citep[see, e.g. ][]{NL:02, GH:05},
and the visible BL photospheres could have much greater emitting areas in comparison with the expected BL model sizes.
This would lead to significant reduction of the averaged BL effective temperature while saving the bolometric luminosity.
This hypothesis can explain the observed properties of the BL in U Gem.
We suggest to call this kind of BLs  "photospheric radius expanded" (PRE) BLs. 
A detailed BL wind investigation would be necessary to test these
ideas.

\section{Conclusions}
 
In this paper we presented the first attempt to compute the  soft X-ray / EUV 
spectra of optically thick boundary-layer models in cataclysmic variables using
the model stellar-atmosphere method. We used the 1D hydrodynamic BL models
calculated by \citet{HKSW:13} for three WD masses (0.8, 1, and 1.2 $M_{\odot}$)
and various values of relative angular velocity (0, 0.2, 0.4, 0.6, and 0.8
$\omega_{\rm K}$). Two other parameters, the mass-accretion rate $\dot M = 1.5
\times 10^{-8} M_{\odot}$\,yr$^{-1}$ and the viscosity parameter $\alpha$ = 0.01
were fixed. Every BL model is divided into a few rings, and every ring model
along the $z$-direction is computed using the model-atmosphere method. The total BL
model spectra are summed over the ring spectra with Doppler effect taken into
account.

The effective temperatures of the considered BL model rings range from 100~kK to
400~kK. At these conditions hydrogen and helium are almost fully ionized and
highly-charged ions of heavy elements such as carbon, oxygen, neon, magnesium,
and silicon determine the opacities and the shape of the model emergent
spectra. { The local ring spectra with numerous absorption lines of these chemical elements together with
absorption edges are smeared by the fast ring rotation and the final model BL spectra show
relatively broad  absorption- and emission-like features, which cannot be identified
with any individual absorption or emission lines. 
The absorption-like features
arise at the spectral parts with high
number densities of strong absorption lines and/or near blue sides of absorption edges. The emission-like features 
arise at spectral parts with a low number densities of absorption lines and/or near red sides of absorption edges. 
An additional smoothing of the spectra due to 
 a finite spectral resolution of X-ray instruments has to be also considered to form these
prominent features. The predominance of spectral lines in the BL
opacities leads to strong BL line-driven winds, which manifest themselves in EUV
observations. We developed a simple method to treat approximately the influence
of a line-driven wind on the BL model structure. This approach is not
self-consistent and it is not sufficiently correct for an investigation of the
wind properties, but it offers the possibility to compute the model emergent
spectra.

The observed soft X-ray and EUV spectra of the dwarf novae SS~Cyg and U~Gem in
outburst are very similar to the computed model BL spectra and exhibit numerous
emission- and absorption-like features \citep{Mauche04, Longetal:96}. Moreover,
we reveal that the soft X-ray {\sl Chandra} spectrum of SS~Cyg in outburst can
be fitted satisfactorily by the spectrum of our BL models  with $M_{\rm WD} =
M_{\odot}$, $\omega_{\rm WD} = 0.8\,\omega_{\rm K}$, and 
$M_{\rm WD} = 0.8\, M_{\odot}$, $\omega_{\rm WD} = 0.6\,\omega_{\rm K}$.}
{with the interstellar absorption parameter $N_{\rm H} \approx 7 \times 10^{19}$\, cm$^{-2}$,
and the distances $\approx$\, 107 and 170 pc correspondingly.   The ratio of the
observed BL luminosity to the disk luminosity of SS~Cyg in outburst is
consistent with the model parameters.  The subtracted spectrum is a spectrum 
of a hot optically thin plasma spectrum and could be associated with the outflow
from the BL surface with the mass loss rate $\sim 10^{17} - 10^{18}$\, g s$^{-1}$,
which is comparable with the assumed mass accretion rate.
Therefore,  future BL models have to be computed with the mass loss taken into consideration. 

On the other hand, the observed EUV flux properties of U~Gem in outburst are
 contradicting  our model predictions.  We suggest that a powerful radiation-driven wind 
from U~Gem's BL could form an extended photospheric BL surface reducing the averaged 
effective temperature but saving the bolometric luminosity. The properties of this
"photospheric radius expanded" BL could be very similar to the observed U Gem BL 
properties. Additional extended BL radiation-driven wind investigations are necessary to
proof this hypothesis. 
 
\begin{acknowledgements} 
We thank Chris Mauche, who put the {\sl Chandra} X-ray spectrum of SS Cyg at our disposal and who helped us to fit it by our models, and the anonymous referee for numerous helpful remarks.
VS thanks DFG for financial support (grant  SFB/Transregio 7 ``Gravitational
Wave Astronomy'')  and Russian Foundation of Fundamental Research (grant
12-02-97006-r-povolzhe-a). MH received financial support from the
German National Academic Foundation (Studienstiftung des deutschen Volkes).
\end{acknowledgements}

\bibliographystyle{aa}
\bibliography{bl}

\end{document}